# Thickness-Dependence of Exciton-Exciton Annihilation in Halide Perovskite Nanoplatelets

*Moritz Gramlich[1,2,*,‡], Bernhard J. Bohn[2,‡], Yu Tong[2], Lakshminarayana Polavarapu[2],*

*Jochen Feldmann[2,*], Alexander S. Urban[1,*]*

[1]Nanospectroscopy Group, Nano-Institute Munich, Department of Physics,

Ludwig-Maximilians-Universität, Munich, Germany

[2]Chair for Photonics and Optoelectronics, Nano-Institute Munich, Department of Physics,

Ludwig-Maximilians-Universität, Munich, Germany

Corresponding Authors

* m.gramlich@physik.uni-muenchen.de (M.G.)
* feldmann@lmu.de (J.F.)
* urban@lmu.de (A.S.U.)

ABSTRACT

Exciton-exciton annihilation (EEA) and Auger recombination are detrimental processes occurring in semiconductor optoelectronic devices at high carrier densities. Despite constituting one of the main obstacles for realizing lasing in semiconductor nanocrystals (NCs), the dependencies on NC size are not fully understood, especially for those with both weakly and strongly confined dimensions. Here, we use differential transmission spectroscopy to investigate the dependence of EEA on the physical dimensions of thickness-controlled 2D halide perovskite nanoplatelets (NPls). We find the EEA lifetimes to be extremely short on the order of 7-60 ps. Moreover, they are strongly determined by the NPl thickness with a power law dependence according to $\tau_2 \propto d^{5.3}$. Additional measurements show that the EEA lifetimes also increase for NPls with larger lateral dimensions. These results show that a precise control of the physical dimensions is critical for deciphering the fundamental laws governing the process especially in 1D and 2D NCs.

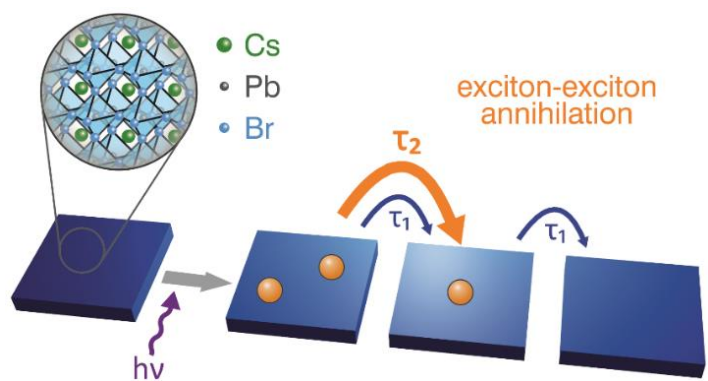

Lead halide perovskite (LHP) nanocrystals (NCs) are an attractive material for realizing highly efficient optoelectronic applications such as light-emitting diodes (LEDs) and lasers.[1-3] LHP-NC-based devices have shown considerable improvement, with LEDs exhibiting more than 20 % EQE in the red and green spectral ranges with less than five years of development.[4-7] However, in order to realize commercialization, the brightness of these devices needs to be increased significantly. This means high pumping (either optically or electrically), which leads to higher carrier densities and promotes nonlinear processes such as Auger recombination or exciton-exciton annihilation (EEA). These processes are detrimental to device functionality as they strongly reduce carrier density, thereby limiting emission intensity and efficiency.[8-10] It is thus essential to understand these multi-particle processes in detail, in order to be able to mitigate their effects. Many of the initial studies focused on quantum dots, wherein a linear dependence of the EEA lifetime (also known as biexciton or bimolecular Auger lifetime[11-13]) on the volume of the nanocrystal was found. This effect was dubbed the "universal volume scaling law" and is independent of the material. [14-16] More recent studies have confirmed the general validity of the volume scaling in perovskite-based systems, albeit with notable differences dependent on the compositions.[17-18] Importantly, for large NCs a sublinear scaling of the EEA lifetimes was observed and attributed to the difference in Coulomb interactions in the weak confinement regime. The strength of confinement accordingly plays an important role in determining EEA lifetimes. This should be especially pronounced for NCs with reduced dimensionality, for example 2D nanoplatelets (NPls) or 1D nanorods (NRs). Such effects were observed both in CdSe- and perovskite-based NCs, however there is significant disagreement as to how the strongly and weakly or non-confined dimensions affect the recombination rates.[19-22] A recent theoretical study on QDs and NRs concluded that it is imperative to include electron-hole correlations to obtain quantitatively accurate lifetimes.[23] Accordingly, it is important to look at these dimensions individually to understand whether and how the degree of confinement effects EEA in NCs especially in halide perovskites and to determine whether the dependencies of EEA are material-independent.

Therefore, in this work, we study the recombination dynamics of excited electron-hole pairs, focusing on the dependence of EEA lifetimes on the thickness of strongly quantum confined LHP NPls. Dispersed in organic solvents, we can assume the NPls to be isolated systems, wherein excitons are created solely through optical excitation with ultrashort laser pulses. Using differential transmission spectroscopy (DTS), we determine the average exciton density per NPl $\langle N \rangle$ as a function of the excitation density $I_{ex}$ of the laser. This correlation is then used to estimate the absorption cross-section $\sigma_{abs}$ at the excitation wavelength for NPls of different thicknesses, $d$. To study EEA, we extract the associated lifetime (denoted as $\tau_2$) in NPls with thicknesses between 2 and 6 ML excited with exactly two excitons by tuning $\langle N \rangle$ precisely. We find the EEA lifetimes to be on the order of tens of picoseconds while exhibiting a nonlinear thickness-dependence with $\tau_2 \propto d^{5.3 \pm 0.2}$. This represents a clear aberration from the "universal volume scaling law", as in the previous reports on CdSe NPls and halide perovskite NCs.[14-15, 17, 19-20, 24-25] Additionally, EEA lifetimes in laterally larger NPls seem to be significantly longer, suggesting that both weakly-confined and strongly-confined dimensions affect the EEA lifetimes. These results give important insights into the nature of EEA or multiexcitonic effects in reduced-dimensionality NCs important for realizing high efficiency and high-power optical lighting or lasing.

The $CsPbBr_3$ NPls investigated here were taken from the same batch synthesized for our original study.[26] They possess a predetermined number of MLs between 2 and 6, corresponding to a thickness $d \approx 1.2 - 3.6\ nm$. The edge length of the square-shaped NPls is 14 ± 4 nm and independent of their thickness. A complete characterization of the NPls (including PL, UV-Vis and TEM analysis) can be found in our previous study by Bohn et al.[26] For all measurements, NPls of the same thickness were dispersed in hexane and diluted to yield an optical density of 0.2 at the laser excitation wavelength of 400 nm. In all measurements, a large ensemble of these NPls is excited simultaneously. Importantly, due to the large spacing, the photogenerated excitons from different NPls cannot interact with each other. Accordingly, every NPl can be treated as an independent system. This is in stark contrast to large bulk crystals, wherein all charge carriers can interact with each other and a rate equation of the form $dn/dt = -n^2 k_2$ governs the nonlinear recombination process.[14]

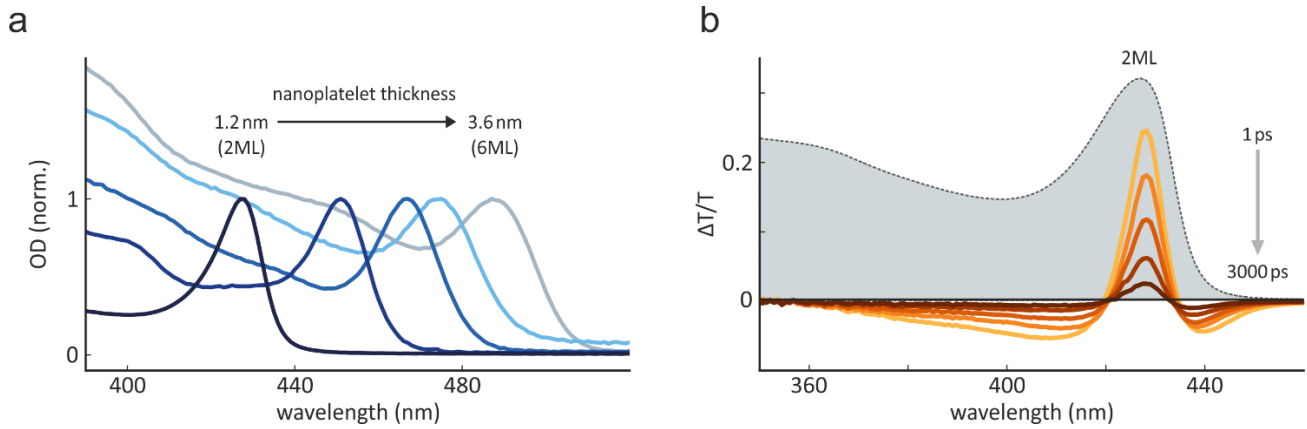

**Figure 1.** (a) Absorption spectra of NPls with thickness between 2 ML and 6 ML normalized to the excitonic peak. A blue-shift and enhancement of the excitonic peak with decreasing NPl thickness are clearly discernible. (b) Differential transmission signal of a 2 ML NPl dispersion excited at 400 nm at times 1-3000 ps after excitation (orange lines). Superimposed in gray is the corresponding linear absorption spectrum, showing that the main $\Delta T/T$ signal occurs at the energetic position of the 1s exciton.

To understand the NPls and any dynamics occurring therein, it is helpful to first consider their linear absorption spectra, as depicted in Figure 1a. Herein, we observe a strong blue shift of the absorption onset with decreasing NPl thickness and a concomitant rise of a more pronounced excitonic peak caused by an increase in exciton binding energy. The obvious thickness-dependence of the spectra is due to quantum confinement induced by the strongly diminished thickness of the NPls. To gain insight into the fast exciton-exciton interaction, ultrafast DTS was applied (see Methods section for more details). With the 100 fs long excitation pulses lying energetically well above the absorption onset, predominantly free electrons and holes are created by the incident photons, yet the predominant DTS signal which builds up within the first

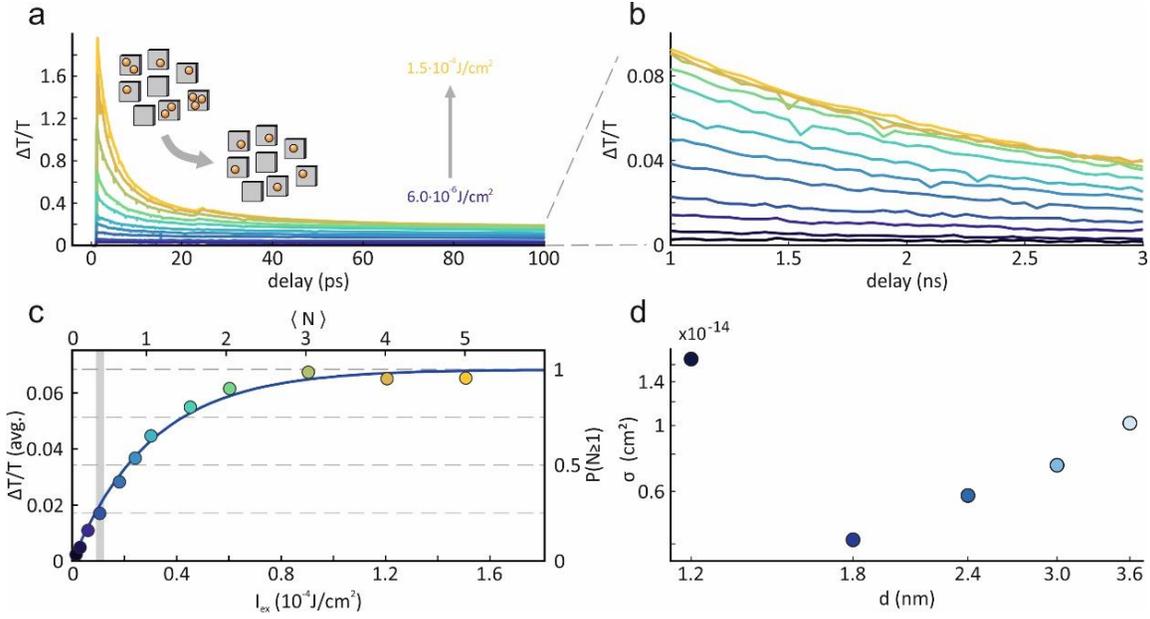

**Figure 2.** (a) Decay of the DTS signal of the 2 ML NPls (from 0-100 ps) for different excitation densities $I_{ex}$. The fast exciton-exciton annihilation process is dominant at higher $I_{ex}$ in the first tens of picoseconds. Inset: Scheme depicting the rapid exciton-exciton annihilation process after which all initially photoexcited NPls contain one exciton while all others do not have excitons. (b) DTS signal from (a) for long time delays (1-3 ns). Above a certain $I_{ex}$, the bleaching signal saturates, meaning that almost every NPl is initially excited with at least one exciton. (c) Saturation of the $\Delta T/T$ signal of the 2 ML NPls after the fast initial process fitted with Poisson statistics. The bleaching value used is the averaged signal shown in (b). The gray bar denotes the range where $0.3 < \langle N \rangle < 0.4$ and which was chosen for the measurements of the exciton-exciton annihilation lifetimes. (d) Absorption cross-sections $\sigma_{abs}$ of the 2 ML to 6 ML NPls calculated from the Poisson statistics via Equation 3

picosecond upon excitation is observed at the location of the 1s exciton, as shown exemplarily for 2 ML NPls (Figure 1b). This behavior was observed for all thicknesses and confirms the fast carrier cooling rates reported for such LHP NPls.[26-27]

The intensity of the DTS signal is a measure of the exciton population in the ensemble.[26] The temporal decay of this population for the 2 ML NPls is plotted in Figure 2a and 2b for a series of different $I_{ex}$. All decay curves exhibit two distinct temporal components. We observe a fast decay component within the first tens of picoseconds. Additionally, this decay component shows a strong $I_{ex}$ dependence and vanishes completely at very low $I_{ex}$. The PL decay at short times can be modelled with a rate equation of $dN/dt = -N^2 k_2$, confirming the bimolecular nature of the exciton-exciton annihilation (EEA) process (see Supporting Information, Figure S1). The second decay component is significantly slower (on the order of nanoseconds), does not show any excitation dependence and can be reproduced through a simple exponential decay function (see Figure S2a).[13] This constitutes the monomolecular recombination of excitons, comprising radiative and nonradiative processes and exhibits an increasing lifetime with increasing NPl thickness (see Figure S2b). With such vastly different timescales, it is safe to assume that after some tens of picoseconds ($\tau_2 \ll t \ll \tau_1$) EEA is essentially over and all initially excited NPls contain a single exciton.[14, 21] From thereon, only monomolecular recombination of single excitons is possible.

To investigate EEA, one must know exactly how many excitons are being created in each NPl. Here we use the fact that excitation is governed by Poisson statistics and saturates at intermediate times ($\tau_2 \ll t \ll \tau_1$) for the highest excitation densities (Figure 2b). We obtain a good estimate of the saturation by averaging the ΔT/T signal in a larger time range (we use 1 - 3 ns) and plotting it in dependence of $I_{ex}$ (Figure 2c, colored dots). The fraction of NPls containing an exciton in this time range given by:

$$P(\langle N \rangle, N \geq 1) = 1 - P(\langle N \rangle, N = 0) = 1 - e^{-\langle N \rangle} \qquad (2)$$

We can fit this formula to the DTS data points and are able to equate the average number of excitons per NPl $\langle N \rangle$ to the excitation density $I_{ex}$. This relation yields the absorption cross section $\sigma_{abs}$ for the NPls (see Supporting Information for details). A detailed derivation of the equation and the subsequential calculation of $\sigma_{abs}$ can be found in the SI and in literature.[13-14] Figure 2d shows the obtained cross-sections at an excitation wavelength of 400 nm for the different NPl thicknesses. All of the values are on the same order of magnitude as the values reported for nanocrystals of similar dimensions.[28] The plot reveals an increase of $\sigma_{abs}$ with increasing thickness from 3 ML upwards but also an even higher σ for the 2 ML sample. The increase of $\sigma_{abs}$, observed from 3 ML upwards, can be explained by two arguments. Firstly, the volume of a NPl, in which a photon can be absorbed, scales with thickness. Secondly, the bandgap decreases with increasing NPl thickness and, therefore, the incident photons with $\hbar\omega = 3.1\ eV$ create excitations higher in the continuum in thicker NPls, where the density of states and consequently the absorption are larger.

Moreover, the obtained value of $\sigma_{abs}$ allows for an estimation of the NPl concentration in the dispersions. To validate our initial assumption that an interaction between the NPls is highly unlikely, we calculated the NPl concentrations in cuvettes with a path length d and an OD of 0.2 at 400 nm using:

$$c = \frac{\ln(10) \cdot 10^3 \cdot OD}{N_A \cdot \sigma \cdot d} \quad (5)$$

For the 2 ML NPls we obtain $c = 1.4 \cdot 10^{14}\ cm^{-3}$, meaning the NPls are separated by an average distance of $200\ nm$. This supports our initial assumption that the interaction between the NPls is negligible in these dispersions, especially on the fast timescale of EEA, which is investigated in the following section. Importantly, this all-optical approach constitutes a convenient method for estimating NC densities in dispersion. Typically, this is done with methods such as inductively coupled plasma mass spectroscopy (ICP-MS) or ICP optical emission spectroscopy (ICP-OES). These require extensive purification procedures, which for perovskite NCs are difficult at best. Consequently, residual precursors in solution can vastly distort the results, whereas the all-optical method is unaffected by this.

Knowing now exactly how many excitons are created in each NPl, we can concentrate on quantifying the EEA process. First, we must look only at the short timescale below 100 ps in the measurements. As EEA is concentration-dependent, the lifetimes will vary strongly, depending on the number of excitons in each NPl ($\tau_2 > \tau_3 > \tau_4 > \cdots$). This can be seen when pumping the NPl dispersions at higher excitation densities (see Figure S3). To compare the annihilation process in NPls of different thicknesses, we extract only the lifetime of an exciton pair in a doubly excited NPl, *i.e.* $\tau_2$. As shown in Figure S4, we can determine the likelihood to find exactly $N$ excitons in a NPl depending on the average number of excitons per NPl $\langle N \rangle$. To minimize the yield of NPls containing more than two excitons upon excitation, while retaining enough signal for analysis, we use excitation densities with $\langle N \rangle \approx 0.3 - 0.4$ (gray shaded areas in Figure 2c and in Figure S4). In this excitation regime, the fraction of doubly excited NPls to all NPls containing multiple excitons is at minimum $\frac{P(N=2)}{P(N\geq 2)} \geq 0.87$. This should be high enough to ensure most annihilation processes occur between only two excitons.

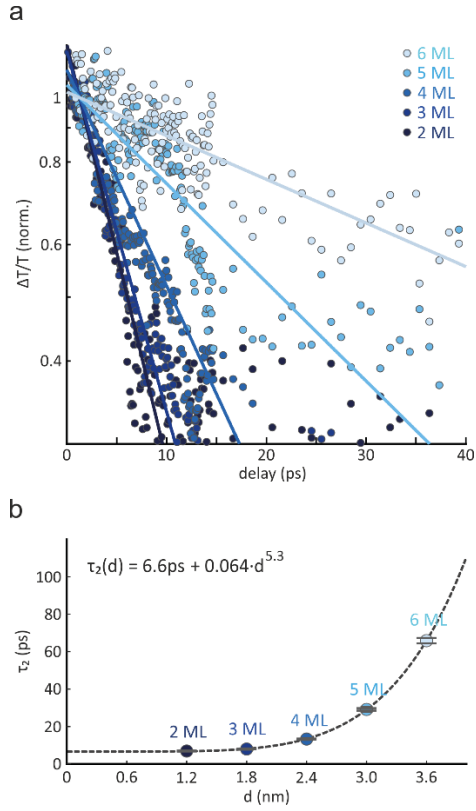

**Figure 3.** (a) DTS decay curves of 2 ML to 6 ML NPls with $I_{ex}$ set to ensure $0.3 < \langle N \rangle < 0.4$. (b) The extracted 1/e lifetimes show a power-law dependence on the NPl thickness $d$, with $\tau_2 \propto d^{5.3}$. The error bars represent the 95% confidence bounds of the fits in (a).

DTS was carried out with these settings and for all NPl thicknesses we extracted the signal at the spectral position of the 1s exciton transition for short times ($t \ll 50\ ps$). The resulting DTS transients exhibit very fast decays on the order of tens of picoseconds for all samples (Figure 3a). We extract the time at which $\Delta T/T$ has dropped to $1/e$ of its initial value and use this as the EEA lifetime $\tau_2$. The extracted values of $\tau_2$ are significantly shorter than in other semiconductor NCs likely due to stronger electron-hole Coulomb interactions.[16-17, 29] Interestingly, they seem to be as long or longer than those previously measured for 5 ML perovskite NPls despite their small lateral size.[20] A clear trend is observable, as the decay becomes progressively faster with decreasing thickness of the NPls. Plotting the EEA lifetime against the NPl thickness $d$, we find that the data matches a simple power dependence law $\tau_2 \propto d^x$ with an exponent of $x = 5.3 \pm 0.2$ (Figure 3b). With their lateral sizes being equal, the volume of the NPls scales linearly with their thickness. Accordingly, this dependence is a clear deviation of the "universal volume scaling law". To our knowledge, the only other report on thickness-dependence of EEA in 2D NPls, showed a similar deviation for the CdSe material system.[19] Therein the authors found a relationship of $\tau_2 \sim d^7$ and developed a theoretical formulism to obtain the EEA lifetimes. They suggested that the weakly and strongly confined dimensions affect this differently, with the former governing the collision frequency of excitons and the latter dimension determining the interaction probability during a collision. We confirmed that the lateral dimension also plays a role, as larger 2 ML NPls (edge length of 35 nm) show longer EEA times (see Figure SI5), also seen by the authors in a subsequent study on LHP NPls.[20] The difference in the exponents between our studies likely stems from the model used to obtain the confinement energy of electrons in NPls.[30] LHPs are very different from typical III-V and II-VI semiconductors, with an s-like conduction band and a p-like valence band as well as similar electron and hole masses. Accordingly, several of the assumptions made in obtaining the quantization energies are not valid for the LHP material system, likely leading to a different dependence of the exciton interaction probability in LHP perovskites. A complex formulism including electron-hole correlations is likely necessary to explain this.[23] Importantly, however, we find that the nonlinear dependence of the EEA lifetimes on NPl thickness is not a material property but a consequence of the specific morphology. Therein, the weakly-confined and strongly-confined dimensions affect the EEA lifetime differently and must be considered individually. A full understanding will likely require a complete synthetic control over thickness and lateral size of the NPls to confirm the theoretical formulism.

In summary, we have investigated recombination dynamics in optically pumped thickness-controlled CsPbBr$_3$ NPls. Using DTS we can estimate the absorption cross-sections of NPls with thicknesses between 2 ML and 6 ML and use these to obtain NPl concentrations in the dispersion. Furthermore, we find extremely fast EEA lifetimes in the NPls consistent with other perovskite NC systems. Most importantly, we find a power law thickness-dependence with an exponent of 5.3 ($\tau_2 \sim d^{5.3}$), constituting a significant deviation from the "universal volume scaling law". Laterally larger NPls also exhibit longer EEA lifetimes. These results show that both weakly- and strongly-confined dimensions play a role in EEA and need to be considered independently, as in previous reports on 1D and 2D NCs. However, LHP NCs possess very different electronic properties than conventional semiconductors, such as CdSe, and a complex theoretical formalism considering electron-hole correlations is likely necessary to postdict the experimentally obtained values. Nevertheless, these results are important both on a fundamental level for the understanding of perovskites and for the realization of high efficiency, high-brightness lighting applications.

**Methods** *NPl synthesis.* The LHP NPls were prepared by exactly following the synthesis route by Bohn et al.[26] In this previous work a detailed analysis of the morphology and the exciton binding energy in these NPls can be found.

*Absorption spectroscopy.* Absorption spectra were recorded using a commercially available Cary 5000 UV-Vis-NIR spectrophotometer by Agilent Technologies, Inc.

*Differential transmission spectroscopy.* The measurements were conducted with a differential transmission spectrometer system by Newport, Inc in combination with a Libra-HE+ Ti:Sapphire amplifier system (Coherent, Inc) for laser pulse generation ($t_{pulse}$ = 100 fs, $f_{rep}$ = 1 kHz, $\lambda_{central}$ = 800 nm). The second harmonic of the laser wavelength was used for excitation of the NPls.

AUTHOR INFORMATION

**ORCID**

Moritz Gramlich: 0000-0002-4733-4708

Bernhard J. Bohn: 0000-0002-0344-7735

Yu Tong: 0000-0002-8828-0718

Lakshminarayana Polavarapu: 0000-0002-9040-5719

Alexander S. Urban: 0000-0001-6168-2509

**Author Contributions**

All authors contributed to writing the manuscript and have given approval to the final version of the manuscript. ‡M.G. and B.J.B. contributed equally.

**Notes**

The authors declare no competing financial interest.

ACKNOWLEDGMENT

We gratefully acknowledge support by the Bavarian State Ministry of Science, Research and Arts through the grant "Solar Technologies go Hybrid (SolTech)" and by the Deutsche Forschungsgemeinsschaft (DFG) under Germany's Excellence Strategy EXC 2089/1- 390776260. This work was also supported by the European Research Council Horizon 2020 through the ERC Grant Agreement PINNACLE (759744). We thank local research clusters and centers (such as CeNS) for providing communicative networking.